\documentclass[aps,prb,twocolumn,showpacs,preprintnumbers,amsmath,amssymb]{revtex4}

\newcommand{\be}{\begin{equation}}
\newcommand{\ee}{\end{equation}}
\newcommand{\bea}{\begin{eqnarray}}
\newcommand{\eea}{\end{eqnarray}}

\usepackage{graphicx}
\usepackage{dcolumn}
\usepackage{bm}
\usepackage{color}
\usepackage{epstopdf}

\newcommand{\sgn}{{\rm sgn}}

\newcommand{\s}{\sigma}

\newcommand{\mycomma}{, }

\begin{document}

\title{The thermal conductance of a two-channel-Kondo model}

\author{ C. P. Moca,$^{1,2}$ A. Roman$^{2}$, D. C. Marinescu$^{3}$ }
\affiliation{
$^1$Institute of Physics\mycomma Budapest University of Technology and Economics\mycomma  H-1521
Budapest\mycomma  Hungary\\
$^2$ Department of Physics\mycomma University of Oradea\mycomma Oradea\mycomma 410087\mycomma Romania\\
$^3$ Department of Physics\mycomma Clemson University\mycomma Clemson\mycomma South Carolina\\
}
\date{\today}

\begin{abstract}
A theory of thermal transport in a two-channel Kondo system,
such as the one formed by a small quantum dot coupled to two
leads and to a larger dot, is formulated.  The interplay of the
two screening constants allows an exploration of the Fermi liquid
and non-Fermi liquid regimes. By using analytical, as well as
numerical renormalization group methods, we study the
temperature dependence of the thermal conductance and the Lorentz number. 
We find that in the low-temperature limit, the Lorentz number attains its
universal value, irrespective of the nature of the ground state.

\end{abstract}


\maketitle

\section{Introduction}
Quantum dots (QD) have long been recognized by
experimentalists and theorists alike for their ability
to embody several important theoretical paradigms.
Some of the most important refer to the possibility
to replicate in a tunable structure the Kondo effect.
First associated with the anomalous resistivity values
in metals, the Kondo effect originates in the antiferromagnetic
interaction of conduction electrons with a local magnetic
moment of spin $1/2$. In this respect, the degenerate ground state
of the electron liquid in the quantum dot functions as a
magnetic impurity, while the electrons in the leads connected to the dot behave as
the surrounding normal metal \cite{ 1-ck}. The multi-channel
Kondo problem involves two or more electron modes in the screening
of the magnetic impurity. The QD representation of this situation has
been realized several years ago and is based on the different interactions
established between an electron in a small dot, that functions as a magnetic
impurity,  with the electrons in two leads and with those located in a
larger dot \cite{Oreg2003}. The relative strength of the screening
realized by these two channels determines the characteristic behavior of the system.

If in the case of a single-channel Kondo (1CK) effect, at temperatures
below the characteristic Kondo temperature, $T_K$, the system behaves
like a Fermi liquid (FL) and is characterized by analytic functions of
temperature, the interplay between the two screenings in the case of
two channel Kondo (2CK) problem, 
permits the realization of a FL, when one screening dominates the other
one, or of a non-Fermi liquid (NFL), when the screenings are equal.
Then each reservoir is trying to screen the magnetic impurity,  
 but in the symmetric limit $J_1 = J_2$, no one succeed. 
This leads to a  NFL type of ground state, where, in contrast 
to the single-channle Kondo problem, the local impurity is only partially screened.
The impact of the Kondo correlations on electronic transport has long been
 an area of active research. Considerably
less is known about their effect on thermal transport, which only recently
has been discussed in several theoretical \cite{Costi2010, Dong2002, Koch2004,
Krawiec2007, Andreev2001, Matveev2002} and experimental \cite{Scheibner2005}
studies that focus on the 1CK model. In this paper we extend such considerations
to the study of the thermal conductivity of a double quantum dot
system that can exhibit a two-channel Kondo state\cite{Potok2006}.  In this picture, the single
electron in the small dot interacts with the electrons in the leads,
with a screening integral $J_1$ and with the electrons in the bigger dot
with a screening integral $J_2$. Besides the Kondo temperature
itself, $T_K$, the relationship between $J_1$ and $J_2$, expressed by $K =
(J_1 - J_2)/J^2$ sets another energy scale of the problem through the characteristic temperature $T^\star = K^2T_K$, smaller than $T_K$.
 For the particular experimental setup realized in Ref. \onlinecite{Potok2006} 
the Kondo temperature ranges between 30 and 130 mK.

\begin{figure}
\includegraphics[width=0.95\columnwidth,clip]{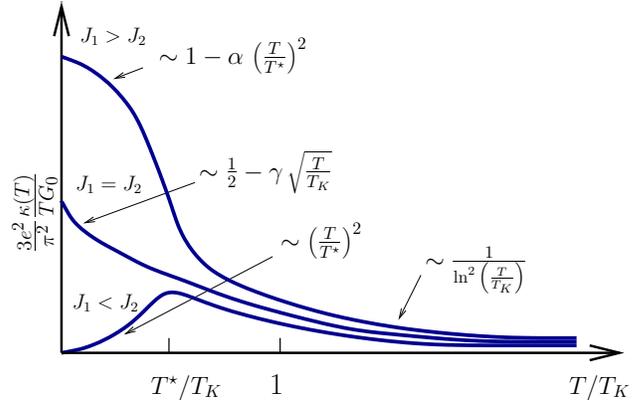}
\caption{\label{fig:sketch} (color online)
The universal behavior of the thermal conductance of a 2CK model
(corresponding to channel 1). Three regimes
are clearly visible when $J_1\ne J_2$.
For $T\ll T^{\star}$ the system is in the Fermi liquid regime,
followed by a crossover ($T^\star<T<T_K$)  to the perturbative ($T\gg T_K$) regime.
The non-Fermi fixed point is realized for $J_1=J_2$. $\alpha$
and $\gamma$ are two numerical constants of order 1, and 
$G_0$ is the universal conductance $G_0 = 2\,e^2/h$.
}
\end{figure}
In an approach that combines analytical arguments with Wilson's numerical
renormalization group (NRG) method, we calculate the thermal conductivity and the
Lorentz number of this system, and investigate  
their behavior as a function of temperature.
In this algorithm, the thermal conductance $\kappa(T)$ corresponding to channel 1 is obtained as
\begin{equation}
\frac{\kappa\left(T\right)}{T}=
\frac{2}{h}
\frac{\pi^2}{3}
\left\{
\begin{array}{l l}
    \theta(K)-\sgn (K)\,\alpha \left( {T /T^\star} \right)^2 & \quad \text{if }K \ne 0\\ \\
    1/2 - \gamma\sqrt{T/T_K}& \quad \text{if }K = 0 \\
\end{array}\right.\;,
\label{eq:thermal_T}
\end{equation} where $\theta(K)$ is the Heaviside function and $\sgn(k)$
is the signum function; $\alpha$ and $\gamma$ are two numerical constants \cite{Adi}
of order $1$. A sketch of this result is presented in Fig. \ref{fig:sketch}.
We find that when $J_1\ne J_2$, at temperature $T <T^*$,
the ground state of the system is a Fermi liquid (FL) and the thermal conductance
varies as $\kappa(T)/T \propto \left( T/T^{\star}\right)^2 $.
The symmetric channel problem, when $J_1 = J_2$, corresponds to a NFL behavior when
the only energy scale in the problem is set by $T_K$. Then, $\kappa(T)/T \propto \sqrt{T/T_K}$.
 In the large temperature regime, when $T\gg T_K$,
a $1/\log^2{(T/T_K)}$ behavior is obtained for $\kappa(T)/T$.

The nature of the ground state strongly affects the
Lorentz number, namely, $L(T) = \kappa(T)/T G(T) $,
with $G(T)$ the charge conductance.
In the Fermi liquid regime, quasiparticles  transport both charge and energy and
the Lorentz number attains a universal value $L(T)= \pi^2k_B^2/3e^2\equiv L_0 $.
This statement is the essence of the Wiedemann-Franz law\cite{Franz1853}, formulated
long time ago for normal metals.
Deviations of $L(T)$ from this behavior are attributed to non-Fermi liquid behavior \cite{Rosch2009}.

The process used in deriving Eq.~(\ref{eq:thermal_T})
adopts the some of the methods that have been previously developed in investigations of
the ground state and the transport properties of the 2CK model
\cite{Pustilnik2004, Toth2007}.

\section{Thermal transport}

The minimal Hamiltonian that
describes the system discussed above is written as
$H = H_0 + H_{\rm int}$ \cite{Affleck1993}. Here,
\be
H_{0} = \sum_{r=\{L,R\},\s} \int d\epsilon\,  \epsilon\,
c_{r,\sigma}^\dagger \left( \epsilon \right)
c_{r,\sigma} \left( \epsilon \right),
\ee 
represents the electrons in the non-interacting leads,
where a single particle state of energy $\epsilon$, spin
$\sigma$ and lead index $r =\{R,L\}$ for left, right
respectively, is associated with the Fermionic
operator $ c_{r,\sigma}^\dagger \left( \epsilon \right)$.
The interacting Hamiltonian
contains two terms, $H_{\rm int} = H_{\rm int}^{(1)}+ H_{\rm int}^{(2)}$,
which describe the coupling of the local spin in the smaller dot  with the
conduction electrons in the external leads
and to those in the second dot. The first term, $H_{\rm int}^{(1)}$, is
\be
H_{\rm int}^{(1)} = {1 \over 2}\,J_1\sum_{r,r' = L,R}\sum_{\s\s'}
\eta_r\eta_{r'}\,{\mathbf S}\, \psi_{r\s}^{\dagger}\,{\boldsymbol\s}_{\s\s'}\,\psi_{r'\s'}.
\ee
Here $\boldsymbol \s$ stands for the Pauli matrices $\boldsymbol \s =\{\s_x, \s_y, \s_z \}$
and $\psi_{r\s}^{\dagger}$'s are the creation field operators, constructed from
the creation operators of the electronic states in the leads,
$\psi_{r\s}^{\dagger} = \sqrt{\varrho} \int_{-D}^{D} d\epsilon\,c_{r,\sigma}^\dagger \left( \epsilon \right)$.
Here $\varrho$ is the conduction band density of states, which
is electron-hole symmetric, $\varrho(\omega) =1/2D$, $-D<\omega<D$.
The coupling of the local spin with the electrons in the leads
is considered to be anisotropic, and will be expressed through some
dimensionless hybridization parameters, $\eta_{L/R}$, of the form,
$\eta_{L/R} = v_{L/R}/(v_L^2+v_R^2)$, with $v_{L/R}$ the amplitude of the hopping
between the dot and the corresponding external lead. An asymmetry
parameter, $\phi$,
allows a more elegant description of the hybridization,
$\eta_L = \cos\left({\phi \over 2}\right)$ and
$\eta_R = \sin\left({\phi \over 2}\right)$.  An even/odd basis emerges,
with new annihilation operators, $\{\Psi, \tilde \Psi \}$, defined as,
$\Psi = \cos\left({\phi \over 2}\right)\psi_{L} + \sin\left({\phi \over 2}\right)\psi_R $ and
$\tilde \Psi  = \sin\left({\phi \over 2}\right)\psi_L - \cos\left({\phi \over 2}\right)\psi_{R}$.
By this unitary transformation, the local spin remains coupled only to the even channel, while the
odd channel becomes decoupled, and then, can be treated as a non-interacting one.

The coupling with the larger dot, 
\be H_{\rm int}^{(2)} = {1 \over 2}\,J_2 \sum_{\s\s'}
{\mathbf S}\, \psi_{2\s}^{\dagger}{\boldsymbol\s}_{\s\s'}\psi_{2\s'}
\ee
is considered to be isotropic, with a dimensionless amplitude $J_2$. The field operators, $\psi_{2\s}^{\dagger}$,
describe the electron-hole excitations in the larger dot.

Our theory permits the calculation
of the thermal conductance of a 2CK system by starting from the heat operator, $Q^{(Q)}$,
\be
Q^{(Q)} = \sum_{\s}\int d\epsilon\, \epsilon \left (c_{L, \s}^{\dagger}(\epsilon) c_{L, \s} (\epsilon)
- (c_{R, \s}^{\dagger}(\epsilon)(c_{R, \s}(\epsilon)   \right ),
\ee
which describes the heat transfer under a temperature gradient between the leads. We implicitly
assume that the leads are in equilibrium, $\mu_L =\mu_R = 0$. The heat current is defined through the usual
expression $I^{(Q)} = i \left [H_{\rm int}^{(1)}, Q^{(Q)} \right ]$.
We introduce the so-called composite fermion
operators,\cite{Costi2000, Zarand2004} ${\cal F}_{\s}^{\dagger}  = \sum_{\s'}\Psi_{\s'}^{\dagger}{\boldsymbol \s_{\s'\s}}\mathbf{S}$, and
$ \tilde h_{\s}^{\dagger} = 2\,\varrho^{3/2}\sqrt{3} \int_{-D}^{D} d\epsilon\,\epsilon\, \tilde c^{\dagger}(\epsilon) $, to rewrite
the heat current as
\be
I^{(Q)} = {1 \over 2\sqrt{3}} J_1\,\sin\left ( \phi \right ) \,\sum_{\s} \left ( {\rm i}\, {\cal F}_{\s}^{\dagger}\, \tilde h_{\s} +{\rm h.c.}  \right).
\label{eq:current_operator}
\ee
The prefactor in the definition of  $\tilde h_{\s} $  was fixed, such that it satisfies the canonical anticommutation relations. Moreover, $\tilde h_{\s} $ describes electrons in the odd-channel. Since this channel is decoupled it may be treated as a non-interacting Fermi system.

When the system approaches the equilibrium, the heat current through the
dot can be calculated with the Kubo formalism.
For that purpose, we consider a temperature gradient being applied between
the external leads. At temperature $T$, the Hamiltonian acquires an additional term,
$H_{T} = Q^{(Q)} \nabla T /T $, with $T$ the temperature itself,
and $\nabla T$ the temperature gradient between the external leads.
Then the average heat current, in its most general form, can be expressed as
\be
\left < I^{(Q)}(t)  \right > = \int_{-\infty}^{t} T\kappa (T, t-t'){ \nabla T \over T}\left (t'\right ) dt',
\label{eq:current_average}
\ee
with $\kappa(T,t)$ denoting the thermal conductance.
While this formalism is general enough to allow the calculation
of the ac-thermal conductance, $\kappa(T, \omega)$, here, we will focus on the dc-limit only.
Algebraic manipulation of Eq.\eqref{eq:current_average} permits expressing
the thermal conductance in terms of heat current operators,
\be
T \kappa (T) = -{\rm i} \int_{0}^{\infty}dt \left < \left [ I^{(Q)} (0), I^{(Q)}(t)  \right ] \right >.
\ee

Finally a simple, analytical form can be derived,
\be
T\kappa (T) = {2 \over h} {\sin^2 \left ( \phi \right )\over 4\sqrt{3} }\sum_{\s}\int d\omega\, \omega^2
\Im m\left \{ { \cal T} (\omega, T) \right \} \frac{\partial f(\omega, T)}{\partial \omega}
\label{eq:thermalconductance}
\ee
In Eq. \eqref{eq:thermalconductance} we can
identify the rescaled Green's function of the
composite operator as the T-matrix, ${\cal T}(\omega)=J_1^2 {\cal G}_{{\cal F}_{\s}}^{R}(\omega)$.

 Eq. \eqref{eq:thermalconductance} allows us
to estimate the thermal conductivity of the 2CK model. We define the Kondo temperature $T_K$ through the
equation, $\Im m {\cal T}(\omega = T_K) = {1\over 2}\Im m {\cal T}(\omega =0)$,
at the symmetric point, $J_1 =J_2$. In this situation,
the screening of the local spin in the smaller dot is equally performed by both channels, their competition leading to an NFL quantum critical state.
Away from the symmetric point, $J_1\ne J_2$,  the
screening is realized by the stronger coupling channel.

In the following, we will always focus on the temperature dependence of the thermal conductivity
for channel 1. In the FL regime, for  $T\ll T^{\star}$,
the T-matrix is analytical \cite{Affleck1993},
\be
\Im m {\cal T}(\omega, T) \simeq
{1 \over \pi \varrho}
 \left(
\theta (K)- \sgn(K) {3\omega^2+\pi^2 T^2 \over T^{\star 2}}
\right).
\ee
At the symmetrical point, and finite temperatures,
the non-Fermi liquid character is manifested by a $\sqrt{T}$ contribution,
\be
\Im m{\cal T} (\omega, T) \simeq {1\over 2\pi\varrho} \left (1 - a\sqrt{\omega/T_K} - b\sqrt{T/T_K}+\dots  \right ) .
\ee
Here $a$ and $b$ are two universal constants.
In this case the T-matrix is purely imaginary and is reduced by
half from the unitary value in
the Fermi liquid regime.
Simple integrations leads to the  qualitative behavior
presented in Fig. \ref{fig:sketch}.

Although this
is a perturbative calculation, in the FL regime,
the ratio of the thermal conductance and the electrical conductance
recovers the Lorentz number exactly in the $T\rightarrow 0$ limit.
It is interesting to point out that this also happens for $J_1=J_2$, when again, for
$\lim_{T\rightarrow 0}$,  $L(T)= L_0$.
This result is somewhat contradictory to the general conclusion that attributes
deviations of the Lorentz number from its constant value $L_0$ to
non-Fermi liquid behavior.
\begin{figure}
\includegraphics[width=0.95\columnwidth,clip]{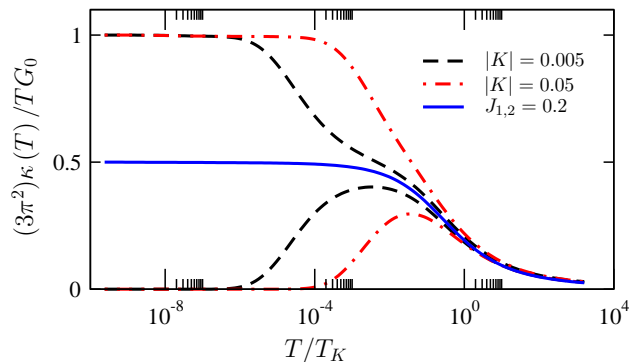}
\caption{
 (color online)
Thermal conductance as function of $T/T_K$ on a logarithmic scale. The NFL fixed point
result, $J_1 = J_2$, corresponds to the solid line, while dashed and dash-dotted
lines correspond to $J_1\ne J_2$. $K$ is the anisotropy factor $(J_1-J_2)/J^2$.
 In both cases the average coupling is $J = (J_1+J_2)/2 = 0.2$.
\label{fig:thermal_Tk}
}
\end{figure}
\section{Numerical results}
\begin{figure}
\includegraphics[width=0.95\columnwidth,clip]{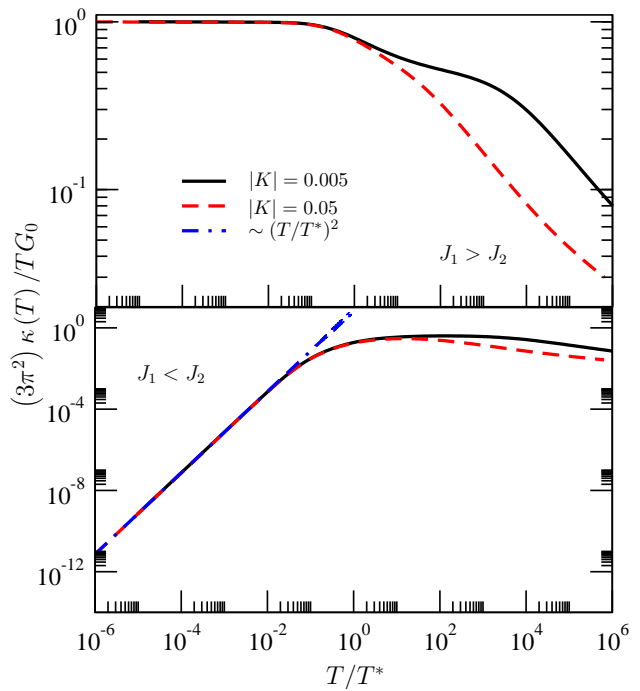}
\caption{\label{fig:thermal_T_star} (color online)
The thermal conductance in the Fermi liquid regime for $K > 0$ (upper panel)
and for $K < 0$ (lower panel). $\kappa(T)/T$ given by Eq. \eqref{eq:thermal_T}
is plotted as a function of $T/T^*$ on a scaled temperature axis.
}
\end{figure}
The analytical results obtained in Eq. \eqref{eq:thermal_T} above were subjected to a numerical test
produced with the
 numerical renormalization group (NRG) method. Within
the NRG framework, the spectral function of the
T-matrix is given as a
weighted sum of $\delta$-functions of the form $\Im m{\cal T}(\omega, T)
\sim \sum_i w_i \delta(\omega-\omega_i)$. This expression replaces the
integral over frequency in Eq. \eqref{eq:thermalconductance}.
The numerical results were obtained with
the Flexible-DMNRG code\footnote{We have used the open access  Flexible-DMNRG code,
available at http://www.phy.bme.hu/$\sim$ dmnrg.}, which
allows the explicit use of the symmetries of the Hamiltonian.
In our particular case we built in the
$SU_{c1}(2)\otimes SU_{c2}(2)\otimes SU_s(2) $
with $SU_{ca}(2)$ the charge $SU(2)$ symmetry in channel $a$ and
$SU_s(2)$ the global spin symmetry. For every run we have
kept 1000 multiplets and we fixed $\Lambda=2$.
\begin{figure}
\includegraphics[width=0.95\columnwidth,clip]{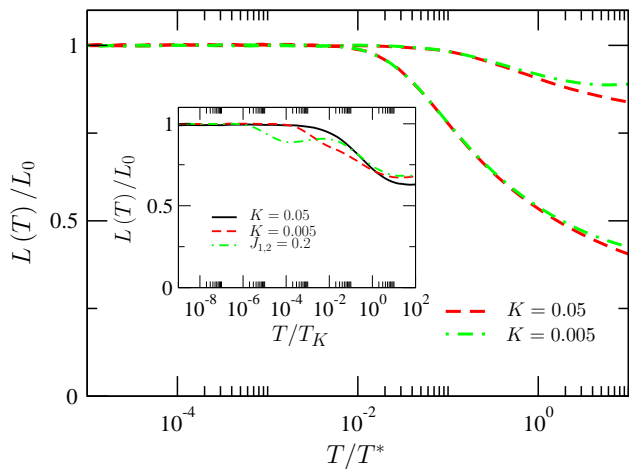}
\caption{\label{fig:lorentz} (color online)
The universal behavior of the Lorentz number as function of $T/T^{\star}$
in the Fermi liquid regime. In the low temperature limit
$T\ll T^{\star}$,  $L(T)$ converges to $ L_0$. The inset presents
$L(T)$ as function of $T/T_K$. At the NFL fixed point, for
$T\ll T_K$ the same universality is recovered.
}
\end{figure}
In Fig. \ref{fig:thermal_Tk} we represent the
results for the thermal conductance $\kappa (T) $
as function of $T/T_K$. At the symmetric point, $J_1=J_2$,
the curves for different $J$'s are scaling one on top of each other and
we have found that the thermal conductance is a universal
function of $T/T_K$, with the behavior predicted by Eq. \eqref{eq:thermal_T}.
In  Fig. \ref{fig:thermal_T_star} we have represented $\kappa(T)/T$
as function of $T/T^{\star}$.  For temperatures $T\ll T^{\star}$,
the universal behavior $\kappa(T)/T \propto (T/T^{\star})^2$, obtained
analytically in  Eq. \eqref{eq:thermal_T}
was also recovered. Thus the numerical results confirm
the universal behavior found analytically.

Finally,
the universality of the Lorentz number was checked numerically.
In Fig. \ref{fig:lorentz}, main panel,
we present the results for the Lorentz number as
function of temperature in the FL regime. The conductance $G(T)$ employed in the calculation was previously shown \cite{Pustilnik2004} to have a universal behavior for $T\ll T^{\star}$, when $G(T) \propto G(T/T^{\star})$. Since both the electric and thermal conductance have a similar behavior,
$L(T) \propto  L (T/T^{\star})$ is found to be a universal function.
At the NFL fixed point, when
$G(T) \simeq G_0 \left (1-\sqrt{\pi\, T/T_K}\right )/2$, the Lorentz number
becomes a universal function of $T/T_K$.
As can be seen in the inset of Fig. \ref{fig:lorentz}, this universal behavior
is valid well below $T_K$, while the scaling in the FL regime goes up
to temperatures as large as $T^{\star}$. In the limit $T\rightarrow 0$, $L(T)/L_0\rightarrow 1$
in all cases.  Similarly, a large violation
of the WF law is observed at large temperatures, $T\gg T_K$, mostly on account of
the suppression of the thermal transport.

In conclusion we have constructed a general framework for the calculation
of the thermal conductance. The method can be easily extended to the computation of
the thermopower, as well as the ac-thermal properties of any
quantum dot system. The NRG method was used to numerically evaluate the conductance, thermal conductance
and the Lorenz number in the 2CK model as function of temperature. Our findings
point towards a universal behavior of all these quantities, similar with the electrical
conductance, studied previously \cite{Pustilnik2004}. In the limit of zero temperature
we have found that $L(T\rightarrow 0)=L_0$ both in the FL and NFL ground states.

This research has been supported by Hungarian grants OTKA No.
K73361, Romanian grant CNCSIS PN II ID-672/2008, the EU-NKTH
GEOMDISS project, and DOE grant number DE-FG02-04ER46139.


\begin{thebibliography}{18}
\expandafter\ifx\csname natexlab\endcsname\relax\def\natexlab#1{#1}\fi
\expandafter\ifx\csname bibnamefont\endcsname\relax
  \def\bibnamefont#1{#1}\fi
\expandafter\ifx\csname bibfnamefont\endcsname\relax
  \def\bibfnamefont#1{#1}\fi
\expandafter\ifx\csname citenamefont\endcsname\relax
  \def\citenamefont#1{#1}\fi
\expandafter\ifx\csname url\endcsname\relax
  \def\url#1{\texttt{#1}}\fi
\expandafter\ifx\csname urlprefix\endcsname\relax\def\urlprefix{URL }\fi
\providecommand{\bibinfo}[2]{#2}
\providecommand{\eprint}[2][]{\url{#2}}

\bibitem[{\citenamefont{Goldhaber-Gordon
  et~al.}(1998)\citenamefont{Goldhaber-Gordon, Shtrikman, Mahalu,
  Abusch-Magder, Meirav, and Kastner}}]{1-ck}
\bibinfo{author}{\bibfnamefont{D.}~\bibnamefont{Goldhaber-Gordon}},
  \bibinfo{author}{\bibfnamefont{H.}~\bibnamefont{Shtrikman}},
  \bibinfo{author}{\bibfnamefont{D.}~\bibnamefont{Mahalu}},
  \bibinfo{author}{\bibfnamefont{D.}~\bibnamefont{Abusch-Magder}},
  \bibinfo{author}{\bibfnamefont{U.}~\bibnamefont{Meirav}}, \bibnamefont{and}
  \bibinfo{author}{\bibfnamefont{M.~A.} \bibnamefont{Kastner}},
  \bibinfo{journal}{Nature (London)} \textbf{\bibinfo{volume}{391}},
  \bibinfo{pages}{156} (\bibinfo{year}{1998}).

\bibitem[{\citenamefont{Oreg and Goldhaber-Gordon}(2003)}]{Oreg2003}
\bibinfo{author}{\bibfnamefont{Y.}~\bibnamefont{Oreg}} \bibnamefont{and}
  \bibinfo{author}{\bibfnamefont{D.}~\bibnamefont{Goldhaber-Gordon}},
  \bibinfo{journal}{Phys. Rev. Lett.} \textbf{\bibinfo{volume}{90}},
  \bibinfo{pages}{136602} (\bibinfo{year}{2003}).

\bibitem[{\citenamefont{Costi and Zlatic}(2010)}]{Costi2010}
\bibinfo{author}{\bibfnamefont{T.~A.} \bibnamefont{Costi}} \bibnamefont{and}
  \bibinfo{author}{\bibfnamefont{V.}~\bibnamefont{Zlatic}},
  \bibinfo{journal}{Phys. Rev. B} \textbf{\bibinfo{volume}{81}},
  \bibinfo{pages}{235127} (\bibinfo{year}{2010}).

\bibitem[{\citenamefont{Dong and Lei}(2002)}]{Dong2002}
\bibinfo{author}{\bibfnamefont{B.}~\bibnamefont{Dong}} \bibnamefont{and}
  \bibinfo{author}{\bibfnamefont{X.~L.} \bibnamefont{Lei}},
  \bibinfo{journal}{J. Phys.: Condens. Matter} \textbf{\bibinfo{volume}{14}},
  \bibinfo{pages}{11747} (\bibinfo{year}{2002}).

\bibitem[{\citenamefont{Koch et~al.}(2004)\citenamefont{Koch, von Oppen, Oreg,
  and Sela}}]{Koch2004}
\bibinfo{author}{\bibfnamefont{J.}~\bibnamefont{Koch}},
  \bibinfo{author}{\bibfnamefont{F.}~\bibnamefont{von Oppen}},
  \bibinfo{author}{\bibfnamefont{Y.}~\bibnamefont{Oreg}}, \bibnamefont{and}
  \bibinfo{author}{\bibfnamefont{E.}~\bibnamefont{Sela}},
  \bibinfo{journal}{Phys. Rev. B} \textbf{\bibinfo{volume}{70}},
  \bibinfo{pages}{195107} (\bibinfo{year}{2004}).

\bibitem[{\citenamefont{Krawiec and Wysokinski}(2007)}]{Krawiec2007}
\bibinfo{author}{\bibfnamefont{M.}~\bibnamefont{Krawiec}} \bibnamefont{and}
  \bibinfo{author}{\bibfnamefont{K.~I.} \bibnamefont{Wysokinski}},
  \bibinfo{journal}{Phys. Rev. B} \textbf{\bibinfo{volume}{75}},
  \bibinfo{pages}{155330} (\bibinfo{year}{2007}).

\bibitem[{\citenamefont{Andreev and Matveev}(2001)}]{Andreev2001}
\bibinfo{author}{\bibfnamefont{A.~V.} \bibnamefont{Andreev}} \bibnamefont{and}
  \bibinfo{author}{\bibfnamefont{K.~A.} \bibnamefont{Matveev}},
  \bibinfo{journal}{Phys. Rev. Lett.} \textbf{\bibinfo{volume}{86}},
  \bibinfo{pages}{280} (\bibinfo{year}{2001}).

\bibitem[{\citenamefont{Matveev and Andreev}(2002)}]{Matveev2002}
\bibinfo{author}{\bibfnamefont{K.~A.} \bibnamefont{Matveev}} \bibnamefont{and}
  \bibinfo{author}{\bibfnamefont{A.~V.} \bibnamefont{Andreev}},
  \bibinfo{journal}{Phys. Rev. B} \textbf{\bibinfo{volume}{66}},
  \bibinfo{pages}{045301} (\bibinfo{year}{2002}).

\bibitem[{\citenamefont{Scheibner et~al.}(2005)\citenamefont{Scheibner,
  Buhmann, Reuter, Kiselev, and Molenkamp}}]{Scheibner2005}
\bibinfo{author}{\bibfnamefont{R.}~\bibnamefont{Scheibner}},
  \bibinfo{author}{\bibfnamefont{H.}~\bibnamefont{Buhmann}},
  \bibinfo{author}{\bibfnamefont{D.}~\bibnamefont{Reuter}},
  \bibinfo{author}{\bibfnamefont{M.~N.} \bibnamefont{Kiselev}},
  \bibnamefont{and} \bibinfo{author}{\bibfnamefont{L.~W.}
  \bibnamefont{Molenkamp}}, \bibinfo{journal}{Phys. Rev. Lett.}
  \textbf{\bibinfo{volume}{95}}, \bibinfo{pages}{176602}
  (\bibinfo{year}{2005}).

\bibitem[{\citenamefont{Potok et~al.}(2006)\citenamefont{Potok, Rau, Shtrikman,
  Oreg, and Goldhaber-Gordon}}]{Potok2006}
\bibinfo{author}{\bibfnamefont{R.~M.} \bibnamefont{Potok}},
  \bibinfo{author}{\bibfnamefont{I.~G.} \bibnamefont{Rau}},
  \bibinfo{author}{\bibfnamefont{H.}~\bibnamefont{Shtrikman}},
  \bibinfo{author}{\bibfnamefont{Y.}~\bibnamefont{Oreg}}, \bibnamefont{and}
  \bibinfo{author}{\bibfnamefont{D.}~\bibnamefont{Goldhaber-Gordon}},
  \bibinfo{journal}{Nature} \textbf{\bibinfo{volume}{446}},
  \bibinfo{pages}{167} (\bibinfo{year}{2006}).

\bibitem[{\citenamefont{Roman et~al.}(2010)\citenamefont{Roman, Moca, and
  Marinescu}}]{Adi}
\bibinfo{author}{\bibfnamefont{A.}~\bibnamefont{Roman}},
  \bibinfo{author}{\bibfnamefont{C.~P.} \bibnamefont{Moca}}, \bibnamefont{and}
  \bibinfo{author}{\bibfnamefont{D.~C.} \bibnamefont{Marinescu}},
  \bibinfo{journal}{unpublised}  (\bibinfo{year}{2010}).

\bibitem[{\citenamefont{Franz and Wiedemann}(1853)}]{Franz1853}
\bibinfo{author}{\bibfnamefont{R.}~\bibnamefont{Franz}} \bibnamefont{and}
  \bibinfo{author}{\bibfnamefont{G.}~\bibnamefont{Wiedemann}},
  \bibinfo{journal}{Ann. Phys. (Berlin)} \textbf{\bibinfo{volume}{165}},
  \bibinfo{pages}{497} (\bibinfo{year}{1853}).

\bibitem[{\citenamefont{Garg et~al.}(2009)\citenamefont{Garg, Rasch, Shimshoni,
  and Rosch}}]{Rosch2009}
\bibinfo{author}{\bibfnamefont{A.}~\bibnamefont{Garg}},
  \bibinfo{author}{\bibfnamefont{D.}~\bibnamefont{Rasch}},
  \bibinfo{author}{\bibfnamefont{E.}~\bibnamefont{Shimshoni}},
  \bibnamefont{and} \bibinfo{author}{\bibfnamefont{A.}~\bibnamefont{Rosch}},
  \bibinfo{journal}{Phys. Rev. Lett.} \textbf{\bibinfo{volume}{103}},
  \bibinfo{pages}{096402} (\bibinfo{year}{2009}).

\bibitem[{\citenamefont{Pustilnik et~al.}(2004)\citenamefont{Pustilnik, Borda,
  Glazman, and von Delft}}]{Pustilnik2004}
\bibinfo{author}{\bibfnamefont{M.}~\bibnamefont{Pustilnik}},
  \bibinfo{author}{\bibfnamefont{L.}~\bibnamefont{Borda}},
  \bibinfo{author}{\bibfnamefont{L.}~\bibnamefont{Glazman}}, \bibnamefont{and}
  \bibinfo{author}{\bibfnamefont{J.}~\bibnamefont{von Delft}},
  \bibinfo{journal}{Phys. Rev. B} \textbf{\bibinfo{volume}{69}},
  \bibinfo{pages}{115316} (\bibinfo{year}{2004}).

\bibitem[{\citenamefont{Toth et~al.}(2007)\citenamefont{Toth, Borda, von Delft,
  and Zarand}}]{Toth2007}
\bibinfo{author}{\bibfnamefont{A.~I.} \bibnamefont{Toth}},
  \bibinfo{author}{\bibfnamefont{L.}~\bibnamefont{Borda}},
  \bibinfo{author}{\bibfnamefont{J.}~\bibnamefont{von Delft}},
  \bibnamefont{and} \bibinfo{author}{\bibfnamefont{G.}~\bibnamefont{Zarand}},
  \bibinfo{journal}{Phys. Rev. B} \textbf{\bibinfo{volume}{76}},
  \bibinfo{pages}{155318} (\bibinfo{year}{2007}).

\bibitem[{\citenamefont{Affleck and Ludwig}(1993)}]{Affleck1993}
\bibinfo{author}{\bibfnamefont{I.}~\bibnamefont{Affleck}} \bibnamefont{and}
  \bibinfo{author}{\bibfnamefont{A.~W.~W.} \bibnamefont{Ludwig}},
  \bibinfo{journal}{Phys. Rev. B} \textbf{\bibinfo{volume}{48}},
  \bibinfo{pages}{7297} (\bibinfo{year}{1993}).

\bibitem[{\citenamefont{Costi}(2000)}]{Costi2000}
\bibinfo{author}{\bibfnamefont{T.~A.} \bibnamefont{Costi}},
  \bibinfo{journal}{Phys. Rev. Lett.} \textbf{\bibinfo{volume}{85}},
  \bibinfo{pages}{1504} (\bibinfo{year}{2000}).

\bibitem[{\citenamefont{Zarand et~al.}(2004)\citenamefont{Zarand, Borda, von
  Delft, and Andrei}}]{Zarand2004}
\bibinfo{author}{\bibfnamefont{G.}~\bibnamefont{Zarand}},
  \bibinfo{author}{\bibfnamefont{L.}~\bibnamefont{Borda}},
  \bibinfo{author}{\bibfnamefont{J.}~\bibnamefont{von Delft}},
  \bibnamefont{and} \bibinfo{author}{\bibfnamefont{N.}~\bibnamefont{Andrei}},
  \bibinfo{journal}{Phys. Rev. Lett.} \textbf{\bibinfo{volume}{93}},
  \bibinfo{pages}{107204} (\bibinfo{year}{2004}).

\end{thebibliography}
\end{document}